\def\lfir{\hbox{L$_{\rm FIR}$}}
\def\msun{\hbox{M$_{\odot}$}}
\def\lsun{\hbox{L$_{\odot}$}}
\def\kms{\hbox{km s$^{-1}$}}
\def\lco{\hbox{L'$_{\rm CO}$}}
\def\lfir{\hbox{L$_{\rm FIR}$}}
\def\lir{\hbox{L$_{\rm IR}$}}
\def\mh2{\hbox{M$_{\rm H_2}$}}

\documentclass[useAMS,usenatbib]{mn2e}

\title[SDSS J0025-10:  a (U)LIRG-QSO transition candidate]{SDSS J0025-10 at $z$=0.30: a (U)LIRG to optical QSO transition candidate \thanks{Based on observations carried out with  the Australia Telescope Compact Array}}
\author[Villar-Mart\'\i n et al.]{M. Villar-Mart\'\i n$^{1}$, B. Emonts$^{2,1}$, M. Rodr\'\i guez$^3$, M. P\'erez Torres$^3$, G. Drouart$^{4,5}$\\
$^1$Centro de Astrobiolog\'\i a (INTA-CSIC), Carretera de Ajalvir, km 4, 28850 Torrej\'on de Ardoz, Madrid, Spain.  villarmm@cab.inta-csic.es \\
$^2$CSIRO Astronomy and Space Science, Australia Telescope National Facility, PO Box 76, Epping NSW 1710, Australia\\
$^3$Instituto de Astrof\'\i sica de Andaluc\'\i a (CSIC), Glorieta de la Astronom\'\i a s/n, 18008 Granada, Spain\\
$^4$European Southern Observatory (ESO), Karl Schwarschild Str. 2, D-85748, Garching bei M\"unchen, Germany\\
$^5$Institut d'Astrophysique de Paris (IAP), 98B Boulevard Arago, 75014 Paris, France}

\begin{document}

\date{Accepted 2013 March 25. Received 2013 March 15; in original form 2012 November 15}

\pagerange{\pageref{firstpage}--\pageref{lastpage}} \pubyear{2002}

\maketitle

\label{firstpage}

\begin{abstract}

We have characterized the amount,  spatial distribution and kinematics of the molecular gas in the merging, double nucleus type 2 quasar SDSS J0025-10 at $z=$0.30 
using the CO(1-0) transition, based on data obtained with the Australia Telescope Compact Array (ATCA).  This is one of the scarce examples of quasar host galaxies  where the CO emission has been 
   resolved spatially at any redshift.
  We infer  a molecular gas mass \mh2=(6$\pm$1)$\times$10$^9$ \msun, which is distributed in two main reservoirs separated by $\sim$9 kpc. $\sim$60\% of the gas is in the central region,  associated with the QSO nucleus and/or the intermediate region between the two nuclei. The other 40\% 
 is associated with the northern tidal tail and is therefore unsettled.
  
With its high infrared luminosity \lir=(1.1$\pm$0.3)$\times$10$^{12}$ \lsun,  SDSS J0025-10 is an analogue of local  luminous LIRGs and ULIRGs. On the other hand, the clear evidence for an ongoing major merger of two gas rich progenitors, the high \lir~   dominated by a starburst, the massive reservoir of molecular gas with a large fraction still unsettled, and the quasar activity  are all properties consistent with a transition phase in the (U)LIRG-optical QSO evolutionary scenario. 
We  propose that we are observing the system during a particular transient phase, prior to more advanced mergers where the nuclei have already coalesced. 
  
We argue that a fraction of the molecular gas reservoir is associated with a tidal dwarf galaxy identified in the optical Hubble Space Telescope  image at the tip of the northern tidal tail. The formation of such structures is predicted by simulations of colliding galaxies.
             
\end{abstract}


\section{Introduction}

CO studies of type 1 quasars (QSO1) at different $z$ have provided accumulating evidence that they often contain  an abundant supply of molecular gas. At redshift $z<$0.5, the range of molecular gas masses spans at least two orders of magnitude: some quasars have \mh2$\la$10$^8$ \msun~ significantly lower than the Milky Way (few $\times$ 10$^9$ \msun) and others contain giant reservoirs $>$10$^{10}$ \msun. 
At the high redshift ($z>$2), quasars contain \mh2~$\sim$several$\times$10$^9$ to $\sim$10$^{11}$ \msun~  (e.g. Bertram et al. \citeyear{ber07}, Xia et al. \citeyear{xia12}). As found for other galaxy types,  these works show that quasars with the highest infrared luminosities also contain the largest \mh2.   

The investigation of the molecular gas content of type 2 quasars (QSO2)\footnote{We refer to QSO2 as radio quiet objects, as opposite to narrow line radio galaxies} has been quite active in recent years, although almost invariably focussed on high $z>$2 objects (e.g. Mart\'\i nez Sansigre et al. \citeyear{mar09}) and with scarce spatial information.

Regarding studies of the molecular gas content of quasars in general, the intermediate redshift  range (0.1$\la z \la$1.5) has remained practically unexplored until very recently (Xia et al. \citeyear{xia12}, Krips, Neri \& Cox \citeyear{knc12}, Villar-Mart\'\i n et al. \citeyear{vm13}).  This redshift range  spans $\sim$60\% of the age of the Universe, an epoch of declining cosmic star formation rate (Hopkins \& Beacom \citeyear{hop06}). Results for QSO2 at redshift $z<$0.5 have been published only in the last year (Krips, Neri \& Cox \citeyear{knc12}, Villar-Mart\'\i n et al. \citeyear{vm13}). They suggest that QSO2 often harbor large reservoirs of molecular gas ($\sim$several$\times$10$^9$ \msun), similar to QSO1 of similar IR luminosity.  
 
 Detailed studies of the spatial distribution of the molecular gas in quasar host galaxies is of special interest. This gas is highly sensitive to the different mechanisms at work during galactic evolution. As such, it  retains relic information about the global history of the systems. The spatial redistribution of the molecular gas during the interactions/merger processes is determinant on the triggering of the starburst and (probably) the quasar activities (Bessiere et al. \citeyear{bes12}, Ramos Almeida et al. \citeyear{ram11}),  while powerful negative feedback might be able to counteract this by cleaning and/or destroying the molecular gas reservoirs and so quenching the star formation and quasar activities (Hopkins \& Beacom \citeyear{hop06}).

Such studies are very scarce and have  also mostly focussed on the highest redshift QSO1 (e.g. Carilli et al. \citeyear{car02}). The main reasons have been so far the lack in sensitivity   and the narrow bandwidths of (sub)-mm telescopes as well as the major interest on the most distant
 sources.  Spatially resolved data on QSO in the local and intermediate redshift ($z<$1.5)  universe are also limited in number 
  (e.g. Feruglio et al. \citeyear{fer10}, Aravena et al. \citeyear{ara11},  Papadopoulos et al. \citeyear{pap08}, Krips et al.  \citeyear{kri07},  Staguhn et al. \citeyear{sta04}). 
The results so far indicate  that a large fraction of the molecular gas in low $z$ quasars (as well as in luminous [10$^{11}$~$\le$\lir/\lsun~$<$10$^{12}$] and ultraluminous [10$^{12}$ ~$\le$\lir/\lsun~$<$10$^{13}$]
infrared galaxies, LIRGs  and ULIRGs) is highly concentrated in the galactic centers, most frequently in rotating disks and rings of $\la$few kpc diameter (e.g. Westmoquette et al. \citeyear{wes12},  Bryant \& Scoville \citeyear{bry99}).  At intermediate $z$, some systems have extended and massive reservoirs of molecular gas, associated with companion objects and/or tidal structures (Aravena et al. \citeyear{ara11},  Papadopoulos et al. \citeyear{pap08}).

We present here a  detailed study of the spatial distribution of the molecular gas traced by the CO(1-0) transition in SDSS J002531.46-104022.2 (SDSS J0025-10 hereafter), a  QSO2 at $z$=0.303 selected from   Sloan Digital Sky Survey (SDSS) database. 
The CO(1-0) ($\nu_{\rm rest}$=115.27 GHz) transition  is the least dependent on the excitation conditions of the gas, which
 is crucial for deriving reliable estimates of the total molecular gas content, including the wide-spread, low-density gas that may be sub-thermally 
 excited (e.g.  Papadopoulos et al. \citeyear{pap01}, Carilli et al. \citeyear{car10}). 

 SDSS J0025-10 has special interest since it is undergoing
a strong transformation via a major merger event
and allows to investigate the distribution and dynamics of
the molecular gas during the process. It moreover shows evidence
for phenomena frequently found in quasars which can
have a profound impact on the evolution of their hosts:
mergers/interactions, star formation and nuclear outflows. A complete study of this system requires
the quantification of the molecular gas content and the characterization of its spatial distribution
and kinematics. This is the purpose of this paper.

\subsection{SDSS J0025-10}

Villar-Mart\'\i n et al. (2011a, 2011b, VM11a and VM11b hereafter)
present a detailed study of this system based on deep optical imaging and spectroscopic data obtained with the Faint Object FOcal Reducer and low dispersion Spectrograph    (FORS2) on the Very Large Telescope (VLT). Based on the high [OIII]$\lambda$5007 luminosity (log($\frac{L_{\rm[OIII]}}{\rm L_{\odot}})$=8.73) and other criteria related to the optical emission line ratios and widths, the object was classified as a QSO2 by \cite{zak03}. This QSO2 is  a  member of an interacting system.
It has two nuclei located at $\sim$5 kpc in projection. The  velocity shift $\Delta$V=-20$\pm$20 \kms~ between them implied by the optical
spectrum  (\cite{vm11a}),   is consistent with their relative motion being approximately constrained on the plane of the sky and thus, the apparent separation is similar to the true separation.  
One nucleus hosts the quasar and the other is forming stars actively. Tidal tails stretch to the North and South.
Their predominance in the broad band optical images shows that they are dominated by stellar continuum emission, although line emission from ionized gas is also detected. 

The optical emission line spectra reveal   extended recent star formation in the companion  nucleus (that sometimes we will refer to as {\it nuc2}) and the northern tidal tail. Since only stars with masses of $>$10 \msun~
and lifetimes of $<$20 Myr contribute significantly to the integrated ionizing flux, this sets an upper limit on the age of the most recent burst of star formation at these locations.     Moreover,  compact knots that are visible in the  tidal tail (see Hubble Space Telescope (HST) image in Villar Mart\'\i n et al. \citeyear{vm12}) are reminiscent of star clusters and/or tidal dwarf galaxies.     The presence of a young stellar population ($<$40 Myr) in the quasar nucleus is also  confirmed by fits of the spectral energy distribution (SED) of the optical continuum (Bessiere et al. 2013, in prep.). 

The quasar nucleus hosts an  ionized outflow of uncertain origin (AGN and/or starburst induced), which produces very broad emission line components with full width half maximum FWHM$\sim$1300 km s$^{-1}$ and blueshifted by $\sim$-80  km s$^{-1}$ relative to the systemic redshift of the galaxy (\cite{vm11b}).

We assume
$\Omega_{\Lambda}$=0.7, $\Omega_{\rm M}$=0.3, H$_0$=71 km s$^{-1}$ Mpc$^{-1}$.  At $z=$0.303, 1\arcsec corresponds to 4.45 kpc.

\section[]{ATCA observations and HST archive data}

The observations  were performed during 2-7 August 2012 with the Australia Telescope Compact Array (ATCA), a radio interferometer in Narrabri, Australia. The ATCA was configured in the most compact hybrid H75 array configuration. Two 2 GHz bands with 1 MHz channel resolution were centered on the redshifted frequency of the CO(1-0) line (88.439\,GHz), providing redundancy in case technical issues would occur with one of the bands. This resulted in a velocity coverage of 7000 km\,s$^{-1}$, maximum resolution of 3.5 km\,s$^{-1}$ and primary telescope beam (i.e. effective field-of-view) of 32 arcsec. Observations were done above an elevation of 35$^{\circ}$ and under good weather conditions, with system temperatures ranging between 450 and 800 K (depending on the antenna and the elevation of the source) and typical atmospheric seeing fluctuations $< 150$ $\mu$m \citep[phase de-correlation of $< 10\%$, see][]{mid06}. The total on-source integration time was 17 hours.

The phases and bandpass were calibrated every 7.5 minutes with a short ($\sim$2 min) scan on the nearby bright calibrator PKS\,0003-066 ($S_{\rm 88.4\,GHz} = 1.8$ Jy at 88.4\,GHz), located at 6.4$^{\circ}$ distance from our target source. Atmospheric amplitude variation were calibrated every 30 minutes using a paddle scan, and telescope pointing was updated every hour, or every time the telescope moved $>$20$^{\circ}$ on the sky. For absolute flux calibration, Uranus was observed close to our target source.

The off-line data reduction was done with the MIRIAD software (Sault, Teuben \& Wright \citeyear{sau95}). Bad data (including data with internal interference or shadowing of an antenna, or data taken during weather conditions that introduced significant phase decorrelation) were discarded. PKS\,0003-066 was used for both phase as well as time-dependent bandpass calibration by interpolating the calibration solution obtained every 7.5 minutes (see \citealt{emo11} for details on time-dependent bandpass calibration within MIRIAD). Atmospheric opacity variations were corrected by weighting the data according to their `above atmosphere' (i.e. paddle-corrected) system temperature. Flux calibration was applied using Uranus [ATCA Uranus model version Aug 2012], resulting in an absolute flux calibration accuracy of 20\%. After Fourier transformation, we obtained a data cube with robust weighting +1 \citep{bri95}. The data presented in this paper were binned by 10 channels and subsequently Hanning smoothed to a velocity resolution of 68 km\,s$^{-1}$, yielding a noise level of 0.7 mJy\,beam$^{-1}$\,chan$^{-1}$. The synthesized beam-size of the data is $6.30 \times 4.39$ arcsec$^{2}$ (PA -85.3$^{\circ}$).   Total intensity images were created by summing all the signal (i.e. without setting a noise treshold)  across the velocity ranges in which CO(1-0) was detected. The calculated CO luminosities in this paper have been derived from these total intensity images (see \S\,3.1).

No 88.4\,GHz radio continuum was detected in the data down to a 5$\sigma$ limit of 0.5 mJy. 

The  HST image, which was obtained with the ACS/WFC  (Advanced Camera for Surveys/Wide Field Camera), was retrieved from the HLA (Hubble Legacy Archive; programme identification 10880 and principal investigator H. Scmitt).
The only processing applied was cosmic ray removal. The accuracy of the HST astrometry is
$\sim$0.3" in both coordinates.
The accuracy of the CO astrometry is expected to be significantly smaller \citep[$\la$0.1''; see e.g.][]{pap08}.

\section{Results}

\subsection{SDSS J0025-10 is a (U)LIRG}

We have constrained the infrared (IR) luminosity, \lir ~(inferred from the 8-1000 $\mu$m wavelength range) by fitting the source spectral energy distribution (SED) defined by the Wide-Field Infrared Explorer (WISE; 3.3, 4.6, 11.6, 22.1 $\mu$m) and Infrared Astronomical Satellite (IRAS; 60 and 100 $\mu$m) photometric measurements, as well as the upper limit on the $\sim$3 mm continuum measured with the ATCA data (see Fig.1).
 Optical photometry has not been used, since this band is known to be a complex mixture of stellar and AGN related components (scattered and/or direct AGN light and nebular continuum; e.g. Tadhunter et al. \citeyear{tad11}, Vernet et al. \citeyear{ver01}). 
To build the SED we used the SWIRE template library (Polletta et al. \citeyear{pol08}) which contains 25 templates including ellipticals,  spirals, starbursts, type 2 and type 1 AGN and composite starburst + AGN. 

The  best fit     corresponds to the starburst template IRAS 20551-4250 (Fig.~1, dashed line). We infer  \lir=(1.4$\pm$0.4)$\times$10$^{12}$ \lsun~for this SED. The starburst template NGC 6240, which also produces a reasonable fit, is also shown for illustration (Fig.~1, dot-dashed line) because this LIRG system  is similar in many aspects to SDSS J0025-10 (double nuclei, interacting local LIRG, Tecza et al. \citeyear{tec00}, Tacconi et al. \citeyear{tac99}). For this SED,  \lir=(8.2$\pm$0.2)$\times$10$^{11}$ \lsun.  

With these two SEDs we constrain  the infrared luminosity   \lir = (1.1$\pm$0.3)$\times$10$^{12}$ \lsun. Taking the uncertainties into account, SDSS J0025-10 is in the transition range of \lir~ values between the LIRG and the ULIRG regimes. 

\begin{figure}
\includegraphics{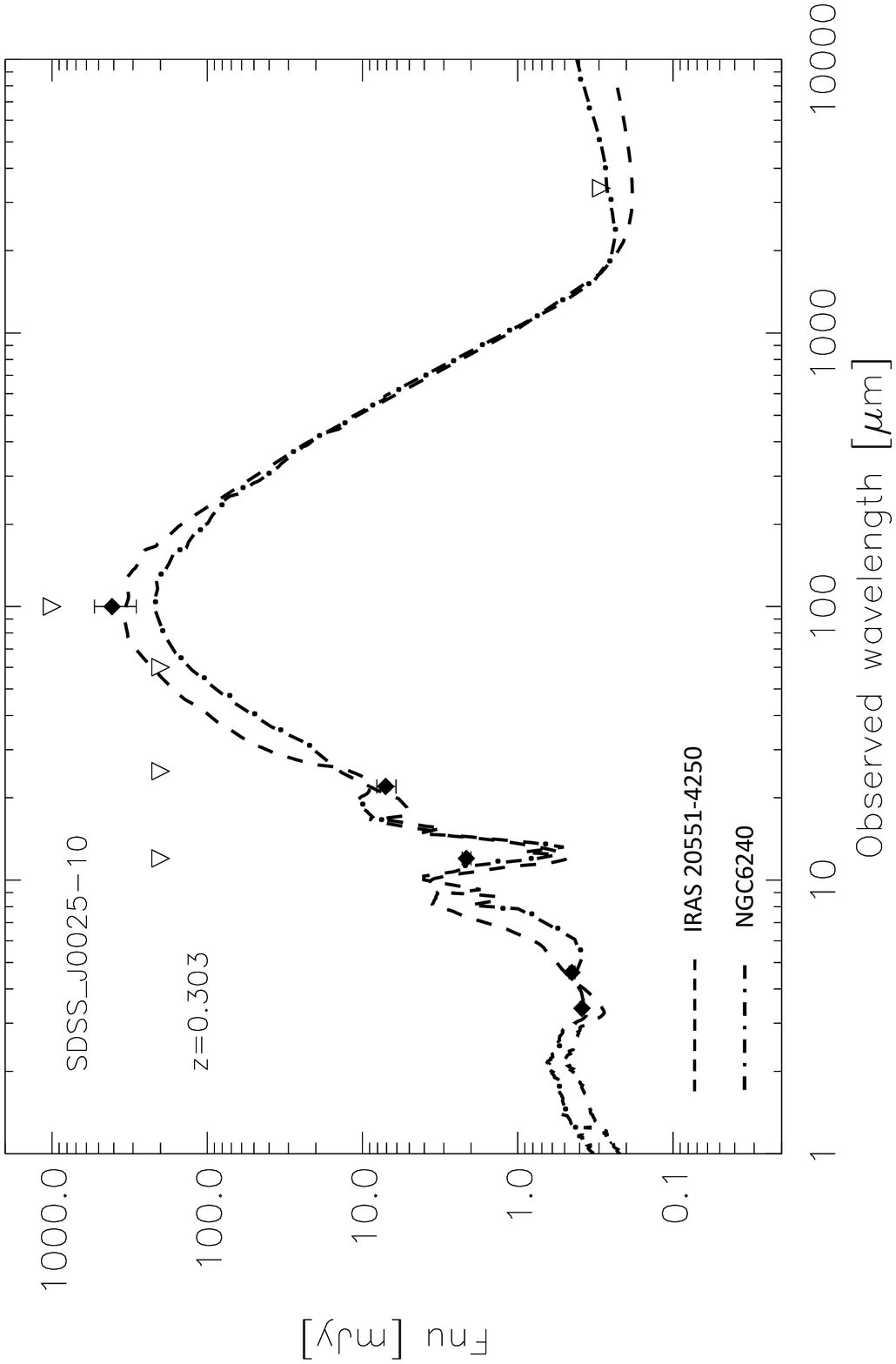}
\vspace{2.5in}
\caption{Fit of the near to far infrared SED of SDSS J0025-10 based on WISE and IRAS photometry and the ATCA 3mm continuum upper limit. Detections and upper limits are marked with diamonds and triangles respectively.  Different  line styles are used for the two SEDs discussed in the text.}
\end{figure}

The fits show that the infrared infrared luminosity of SDSS J0025-10 is dominated by a starburst. We calculate the star forming rate as SFR = 1.73$\times$10$^{-10}$$\times$\lir~  (\msun~yr$^{-1}$) = 190$\pm$52 \msun~yr$^{-1}$, assuming  solar abundances and a Salpeter Initial Mass Function (IMF) (Kennicutt 1998).

\subsection{Mass and spatial distribution of the molecular gas}

\begin{figure}
\includegraphics{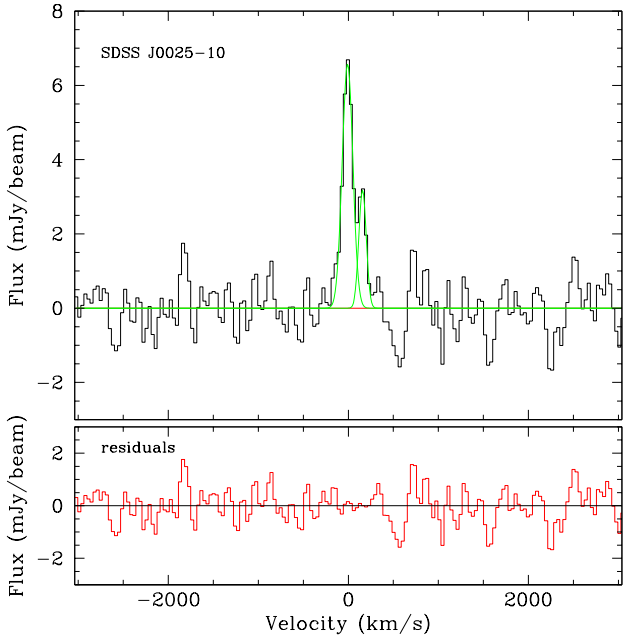}
\vspace{3.4in}
\caption{1-dimensional spectrum of SDSS J0025-10, extracted against the QSO2 nucleus (2.3''$\times$2.3''). Notice the double horned profile, with Gaussian fits visualizing the two components described in the text.  The bottom plot shows the residuals after subtracting the two Gaussian profiles from the spectrum. Velocity values are relative  to the peak of the `blue' component, which we assume as the systemic velocity  ($z$=0.3031$\pm$0.0001, see text).}
\end{figure}

 We measure a total CO(1-0) luminosity in SDSS J0025-10 of \lco=(7.5$\pm$1.5)$\times$10$^9$  K \kms~pc$^2$. $\rm L'_{CO}$ is calculated as:

$$\rm \lco~ = 3.25 ~ \times 10^7 (\frac{S_{CO} \Delta V}{\rm Jy~km/s})~(\frac{D_L}{Mpc})^2 ~(\frac{\nu_{rest}}{GHz})^{-2}(1 + z)^{-1}  $$

where  $\rm I_{CO} = S_ {CO} ~\Delta V$ is  the integrated CO(1-0) line intensity in Jy km s$^{-1}$, $\rm D_L$ is the luminosity distance in Mpc
 and  $\nu_{\rm rest}$=115.27 GHz, is the rest frame frequency of the CO(1-0) transition  (Solomon \& Vanden Bout \citeyear{sol05}). 
 With this \lco, the system follows the \lco~vs. \lir~ (or \lfir) correlation defined by different types of galaxies  (Solomon \& Vanden Bout \citeyear{sol05}), including QSO1 and QSO2 (Bertram et al. \citeyear{ber07}, Villar-Mart\'\i n et al. \citeyear{vm13}, Krips, Neri \& Cox \citeyear{knc12}).
The molecular gas mass is then calculated as \mh2 = $\alpha \times$\lco =(6$\pm$1)$\times$10$^9$ \msun,    assuming $\alpha$=0.8 \lsun~ (K \kms~ pc$^2$)$^{-1}$. 
This  value of $\alpha$ is often used in studies of (U)LIRGs and active galaxies (e.g. Downes \& Solomon \citeyear{dow98}), although a range $\alpha \sim$0.3-1.3 is possible (e.g.  Solomon \& van den Bout \citeyear{sol05}, Sanders \& Mirabel \citeyear{san96}).  Similar amounts of molecular gas have been measured for quasars (both type 1 and type 2) of similar infrared luminosity ~and redshift (e.g. Krips, Neri \& Cox \citeyear{knc12}, Villar-Mart\'\i n et al. \citeyear{vm13}).

The CO(1-0) line shows a double horned profile (Fig.~2) with two kinematic components. The blue component   has  $z$=0. 3031$\pm$0.0001 (which we assume as the systemic redshift $z_{\rm sys}$) and FWHM=140$\pm$25 \kms. The red, fainter component is shifted by +160$\pm$10 \kms~  and has FWHM=80$\pm$40 \kms. A negative depression (around v\,=\,500\,km\,s$^{-1}$) is also visible in Fig.\,2. This feature appears at -2.8$\sigma$ (when integrated over its full extent). Fig.\,2 (bottom) shows that it may be merely a prominent noise peak in the data. Since this negative depression is not noticeably caused by external effects (e.g. weather or phase calibration), it may also indicate that there are low-level systematics affecting the ATCA 3mm (similar to low-level systematics described by \citealt{emo11} that occur at the extreme edge of the ATCA 7mm band, or possible baseline artifacts mentioned by \cite{nor13} in early wide-band 3mm ATCA data). Regardless, the integrated blue and red part of the double horned CO profile are detected at the 10$\sigma$ and 7$\sigma$ significance level respectively and thus unambiguously stand out above the noise, indicating the reliability of the CO emission discussed in this paper.

  \begin{figure*}
\includegraphics{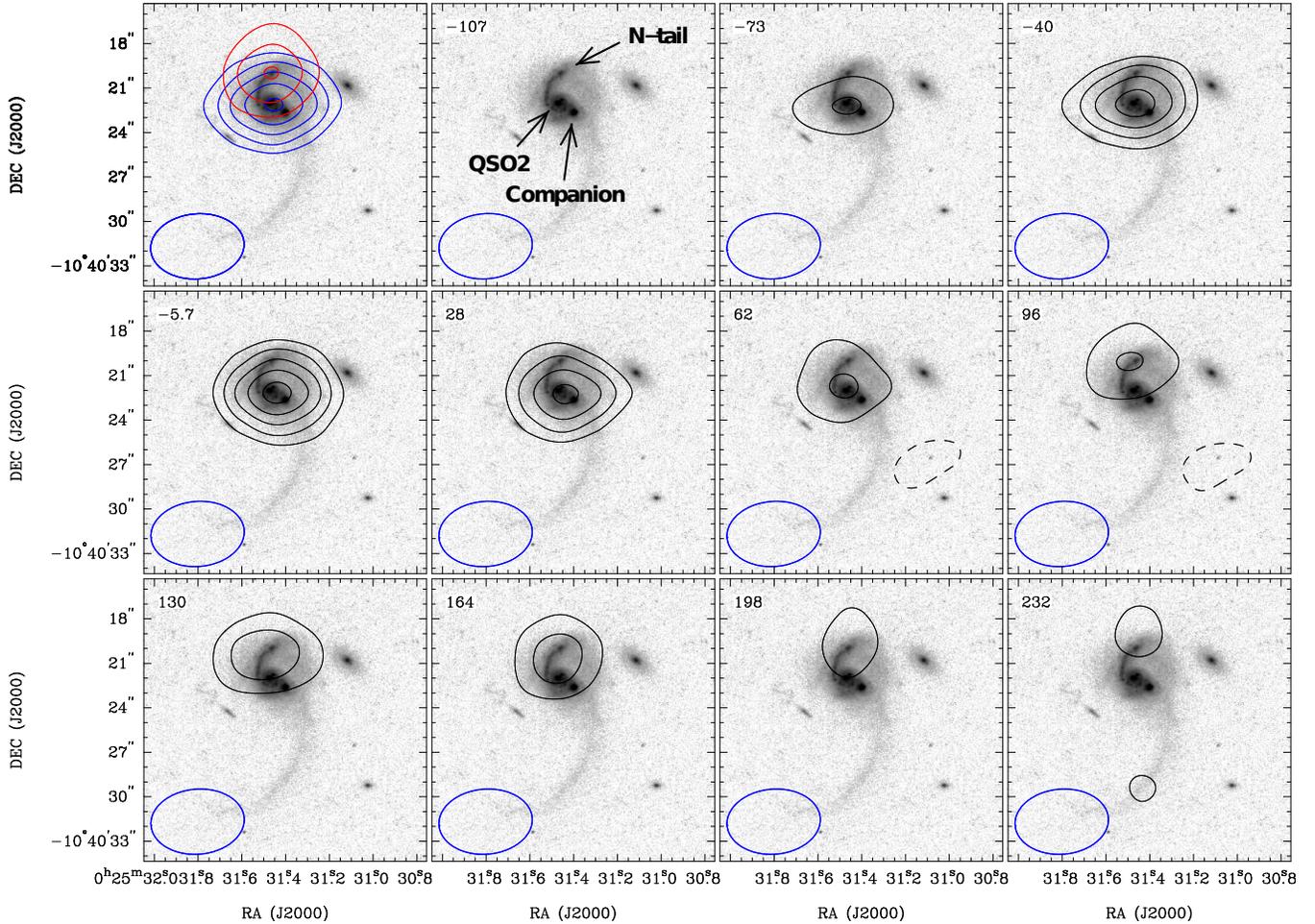}
\vspace{5.3in}
\caption{The total intensity image of both the blue (J0025A, $-90 \leq {\rm v} \leq 79$\,km\,s$^{-1}$) and red (J0025B, $79 \leq {\rm v} \leq 249$\,km\,s$^{-1}$) part of the CO(1-0) signal is shown in the first frame, followed by the channel maps of the CO emission. Contour levels of the total intensity map:  0.27, 0.41, 0.54, 0.68, 0.81 Jy bm$^{-1}$ $\times$ km\,s$^{-1}$; contour levels channel maps: -3.0 (dashed), 3.0, 4.5, 6.0, 7.5, 9.0 (solid) $\times$ $\sigma$, with $\sigma$=0.7 mJy\,bm$^{-1}$. Velocity values are relative to $z_{sys}=0.3031 \pm 0.0001$. The location of the QSO2, companion star forming nucleus  and the northern tidal tail are indicated in the second panel. A color version of this figure can be seen in the electronic form.}
\end{figure*}

The spatial distribution of  CO(1-0) in different channel maps is shown in Fig.~3 overlaid on the HST optical image. The CO(1-0) emission detected at  $\ge$3$\sigma$ level  spans a range of velocities $\sim$[-110,+230] \kms~ relative to $z_{\rm sys}$. The CO(1-0) is spatially extended and accumulated in two main reservoirs  separated by $\sim$2\arcsec or $\sim$9 kpc  (blue and red contours in Fig.~3, first panel) which we will name J0025A and J0025B.  The similarity within the errors of the velocity shift   between them, compared to the two kinematic components isolated on the 1-dimensional spectrum suggests  that
J0025A and J0025B are responsible for the double horned profile discussed above. J0025A and J0025B  have CO luminosities of  (4.3$\pm$0.9)$\times$10$^9$  and (3.1$\pm$0.9)$\times$10$^9$ K \kms pc$^2$ respectively\footnote{Values have been derived from the total intensity image in Fig.3, but were corrected for line contamination and missed low-level flux based on the Gaussian fitting of the profiles (Fig.~2). This did not significantly change the values for the blue component, but increased both the estimated value and the associated uncertainty of the flux of the red component by 10\%. The errors also reflects the 20\% uncertainty in absolute flux calibration (\S2).}, which correspond to 
\mh2=(3.4$\pm$0.7)$\times$10$^9$  and (2.5$\pm$0.7)$\times$10$^9$ \msun. 

In Fig.\,4 we show the  location of the peaks of the CO(1-0) emission in J0025A and J0025B overlaid onto the HST image. These locations have been derived by fitting a Gaussian profile to the integrated CO(1-0) emission. Within the uncertainty, this Gaussian profile could be represented by a point source with the size of the synthesized beam. The locations of the CO(1-0) peaks in the different channel maps of Fig.\,3 are also plotted, including the uncertainty due to fitting and astrometric errors (see caption for details). From Fig.\,4 it is clear that   the  CO(1-0) emission in     J0025B is  associated with the northern tidal tail while J0025A
 is preferentially associated with the QSO2 nucleus and/or the region {\it between} both nuclei. 

 We do not detect any CO(1-0) emission associated with the  companion nucleus {\it nuc2}. However, the star forming activity it habours (\S\,1.1) requires a local supply of molecular gas.  The  H$\beta$ luminosity  from {\it nuc2} (\cite{vm11a})  implies a line reddening corrected SFR$\sim$8 \msun~yr$^{-1}$, assuming SFR(M$_{\odot}$ yr$^{-1}$) = 2.3$\times$10$^{-41}$ $\frac{\rm L_{\rm H\beta}}{\rm erg s^{-1}}$  for solar abundances and a Salpeter IMF (Kennicutt \citeyear{ken98}).\footnote{For comparison, the SFR over the entire Milky  Way is several \msun~yr$^{-1}$ (e. g. Robitaille \& Whitney \citeyear{rob10})} This in turns implies \lir$_{\it nuc2}$$\sim$4.6$\times$10$^{10}$ \lsun, which would be a lower limit if part of the star formation were totally obscured (Rodr\'\i guez Zaur\'\i n et al. \citeyear{rod11}). This is $\sim$8\% of the total \lir~of the system. If {\it nuc2} follows the  \lir~vs. \lco~correlation defined by star forming galaxies (see Fig.~8 in Solomon \& van den Bout \citeyear{sol05}), it roughly implies  that  it  contains $\ga$8\% of the total molecular gas or $\ga$5$\times$10$^8$ \msun. The 3$\sigma$ upper limit for a CO(1-0) non-detection with FWHM = 140 \kms~  in our data is \lco = 1.0$\times$10$^9$ K \kms pc$^2$, or \mh2 = 8$\times$10$^8$ \msun. Although the calculations are highly uncertain due to the large scatter of the \lir~vs \lco~ correlation and the uncertainty on the amount of obscured star formation in {\it nuc2}, they imply that its CO(1-0) emission likely falls below the detection limit of our data (or otherwise a low-level signal would likely be hidden within the large ATCA beam by  the dominant emission from the quasar nucleus and/or intermediate region).
 New observations with higher spatial resolution and sensitivity are required to both determine the exact location of the CO in J0025A and measure (or set reliable limits on) the CO content of the star forming companion nucleus.

 \section{Discussion}

\subsection{SDSS J0025-10 vs. local (U)LIRGs}

SDSS J0025-10 is a (U)LIRG double nuclei merging system with all observed properties consistent with those of  numerous local ($z\la$0.05) luminous LIRGs (several$\times$10$^{11} \la$ \lir $<$10$^{12}$ \lsun)  and ULIRGs. It shows  clear morphological evidence for an ongoing major merger: the double nucleus, the tidal tails,  the inner spiral structure (which surrounds the QSO nucleus clockwise and apparently becomes the northern tidal tail) and  the stellar clusters and/or tidal dwarf galaxies observed along this structure (Fig.~4)   are all morphological features often observed in ULIRGs and  luminous LIRGs (Haan et al. \citeyear{haan11}, Miralles Caballero et al. \citeyear{mir11}, Monreal Ibero et al. \citeyear{mon07}).  As in these systems, the star formation and probably the AGN activity are triggered by the interaction   (see Alonso-Herrero \citeyear{alo13} for a recent review; see also Bessiere et al. \citeyear{bes12}).

The IR luminosity, \lir=(1.1$\pm$0.3)$\times$10$^{12}$ \lsun, which is dominated by the starburst component, the large molecular gas content  \mh2=(6$\pm$1)$\times$10$^9$ \msun,  the SFR=190$\pm$52 \msun~yr$^{-1}$, the star formation efficiency   SFE=$\frac{\rm \lir }{\rm \mh2}$=183$\pm$62 \lsun~ \msun$^{-1}$ and the gas exhaustion time scale $\tau_{\rm SF}$ = $\frac{\rm M_{H_2}}{\rm SFR}$= 32$\pm$11 Myr are   all in
the range  of  luminous LIRGs  and ULIRGs (Solomon \& van den Bout \citeyear{sol05}, Sanders \& Mirabel 1996, Alonso-Herrero \citeyear{alo13}). 

In SDSS J0025-10 one nucleus hosts the AGN and both harbour star formation. 
About 20-30\% of local LIRGs have an optically identified AGN, with this percentage increasing with the \lir ~and becoming $\sim$70\% for ULIRGs  (e.g. Yuan,  Kewley \& Sanders \citeyear{yuan10}, Nardini et al. \citeyear{nar10}, Veilleux et al. \citeyear{vei99}, \citeyear{vei95}). On the other hand, 
a large fraction ($\sim$63 \%)  of local luminous LIRGs and ULIRGs  have double nuclei (Haan et al. \citeyear{haan11}) 
with a variety of scenarios. In some cases both nuclei host star formation and AGN activities (being NGC6240 the most famous example; e.g. Tecza et al. 2000, Tacconi et al. 1999);  in others star  formation is detected in both nuclei, but none seem to harbour AGN activity or show a composite AGN+starburst  spectrum, like SDSS J0025-10 (e.g.  Yuan,  Kewley \& Sanders \citeyear{yuan10}).

The CO gas in SDSS J0025-10 is distributed in two main spatial components, J0025A and J0025B of masses (3.4$\pm$0.7)$\times$10$^9$  and (2.5$\pm$0.7)$\times$10$^9$ \msun~ respectively. They are separated by 2\arcsec or $\sim$9 kpc (projected distance) in space and  160 \kms~ in velocity. J0025A is preferentially associated with the quasar nucleus and/or the intermediate region between the two interacting nuclei, while J0025B is associated with the northern tidal tail. CO(1-0) emission from the companion nucleus is not unambiguously detected, although the active star formation it hosts  implies that there must be a local reservoir of  \mh2$\ga$several$\times$10$^8$ \msun. The CO gas in local (U)LIRGs 
  is usually highly concentrated in the central region towards the core of the merger ($r\la$1 kpc, e.g.  Bryant \& Scoville \citeyear{bry99}, Sanders \& Mirabel \citeyear{san96}).
In addition, a significant fraction of the gas is sometimes spread over scales of 10 kpc or more.  This is possibly the case of SDSS J0025-10:
J0025A might be   the compact, central concentration (although higher spatial resolution observations are  required to constrain its exact location and  compactness), while J0025B is the more extended gaseous component.

\subsection{SDSS J0025-10  as a (U)LIRG-optical QSO transition object}

Both theory and observations support  that the merger of two gas rich progenitor galaxies form a new, more massive elliptical galaxy (e.g. Toomre \& Toomre \citeyear{too72}, Barnes \& Hernquist \citeyear{bar96}). As a consequence of the gravitational interaction, gaseous dissipation funnels gas
 into the the gravitational center and triggers intense star formation, which is manifested in the (U)LIRG phenomenon. The gas  inflow could also fuel the formation of an AGN (e.g. Hopkins et al. \citeyear{hop08}, \citeyear{hop06}).  The discovery of massive reservoirs of molecular gas
 shifted far away from the quasar location has prompted some authors to suggest that  gas poor/gas rich galaxy mergers can also trigger quasar activity (Riechers \citeyear{rie13}, Aravena et al.  \citeyear{ara11}, Papadopoulos et al. \citeyear{pap08}).

 We are observing SDSS J0025-10 when the galaxies are deeply interpenetrating and approaching the final coalescence ($r\la$a few kpc).  Simulations of gas dynamics and starbursts in major gas rich mergers with or without quasar activity (e.g. Mihos \& Hernquist 1996, Hopkins et al. \citeyear{hop05}) show that at this stage   
 the morphologies  show distorted, irregular isophotes and extended tidal tails, as observed in SDSS J0025-10. 
 The models also predict that at this moment 
 the gas experiences the strongest levels of inflow so that large amounts of gas  accumulate in the central region and very intense   star formation activity is triggered as a consequence. The duration of the starburst is expected to be very short $\sim$50 Myr (Mihos \& Hernquist \citeyear{mih94}). The presence of such a young stellar population is confirmed in SDSS J0025-10 (see \S 1.1).

 \cite{san88a} proposed that there is a natural evolution from ULIRG to QSO: ULIRGs are the initial dust-enshrouded stage of quasars, which will become optical quasars once the nuclei shed the obscuring dust via feedback mechanisms (see also Hopkins \& Beacom \citeyear{hop05}).    Some  authors have expanded the ULIRG-QSO evolutionary scenario accommodating luminous LIRGs as a phase that appears earlier  in mergers  than the ULIRG phase  (e.g. Haan et al. \citeyear{haan11}, Yuan, Kewley \& Sanders \citeyear{yuan10}).  A review on the observational and theoretical studies, with pros and cons  of  these scenarios can be found  in the recent paper by \cite{rot13}.  
 Although the   merger driven evolutionary sequence (U)LIRG-QSO  probably is too simplistic (Veilleux et al. \citeyear{vei02}), both theory and observations support that at least a significant fraction of optical quasars might evolve from a prior (U)LIRG phase.

In such scenario,  SDSS J0025-10 is a strong candidate to be a (U)LIRG-optical QSO transition object. All the properties investigated here   are expected for such a transition phase:  high infrared luminosity powered by intense star formation, a rich molecular gas reservoir, strong evidence for a major merger of two gas rich progenitors and  optical identification of quasar activity (Sanders et al. \citeyear{san88a}, \citeyear{san88b}).

 The nuclear separation in SDSS J0025-10 ($\sim$5 kpc) is   consistent with typical values of double nuclei LIRGs and significantly larger  than  in more advanced mergers (e.g. Haan et al. \citeyear{haan11}, Yuan, Kewley \& Sanders \citeyear{yuan10}), which are more frequently found in ULIRGs. This and   the fact that a large fraction of the molecular gas  is still unsettled show that the system is in an transient stage prior to the final coalescence of the nuclei. 
 
 The identification of an ionized outflow  associated with the quasar nucleus (\cite{vm11b}), prompts the interesting possibility of tracing its signature in both the neutral and molecular phases to fully characterize its mass, geometry and energetics. This information would allow to evaluate whether the outflow is powerful enough to   clear up completely the material surrounding the quasar nucleus, quench the star formation activity and convert SDSS J0025-10 into an optical quasar before becoming a non active massive elliptical galaxy
 (e.g. di Matteo, Springel \& Hernquist \citeyear{dim05}).

 \subsection{A large reservoir of molecular gas at the tip of the  tidal tail}
 
 Merger simulations of gas rich galaxies show that, besides the infalling gas which concentrates  in a compact central region of less than 1 kpc in size,  part of the gas survives the merger (Barnes \& Hernquist 1996; Mihos \& Hernquist 1996; Hopkins et al. \citeyear{hop08}). This is primarily material 
which has been moved to large radii temporarily, either blown
out by a combination of supernova and AGN feedback or thrown
out in tidal tails.  The lack of spatial information prevents us from making a confident elucidation of the nature of J0025A and J0025B but given their spatial location, it is  reasonable  to propose that J0025A is  (or will become) the compact component while  J0025B,  associated with the northern tidal tail, is   ``surviving"  material.

 In Fig.~4, it seems  that the locations of the peak of the CO(1-0) emission in the various channel maps  traces the tidal tail, from the inner spiral-like structure between the nuclei to  the tip of the tail, maybe beyond.  This suggests that there is molecular gas along the tail.  According to the model predictions, this surviving gas will then slowly settle back  into a
small, rotationally supported embedded disk  (Hopkins et al. 2008, Bournaud \& Duc \citeyear{bou06})  $\sim$a couple $\times$10$^8$ yr after the coalescence.
 This gas is not expected to contribute much to the galaxy stellar mass ($\sim$a few per cent), even for the most gas rich mergers.
 Thus, although a large reservoir of gas is still falling towards the center, it seems unlikely that it will produce a significant enhancement on the build up of the galaxy stellar mass.

On the  other hand, the centroid of J0025B is coincident with  a prominent feature of elliptical shape located at the tip of the northern tidal tail  and clearly identified in the HST image (Fig.~4). A CO accumulation at a tidal tail's tip is a promising tidal dwarf galaxy (TDG) candidate (Bournaud \& Duc \citeyear{bou06}).  A TDG is a self-gravitating
entity  that was made up from the debris of a
galaxy interaction.

 Indeed, the properties of the optical feature strongly support this scenario. The optical size  along the major axis 
$D=$1.4$\pm$0.1 kpc, the effective radius $R_{eff}=$0.70$\pm$0.06 kpc, the observed H$\beta$ luminosity  L(H$\beta$)=3.6$\times$10$^{40}$ erg s$^{-1}$ (\cite{vm11a}) are all consistent with values measured for TDGs (e.g. Monreal Ibero et al. \citeyear{mon07}, Bournaud \& Duc \citeyear{bou06}).
The J0025B CO(1-0)  line luminosity is $\ga$50 times higher (and therefore implies $\ga$50 times more molecular gas mass) than typically measured for TDGs (e.g. Braine, Duc \& Linsfeld \citeyear{bra01}), but this most probably reflects the CO(1-0) contamination of J0025B by emission from other regions.

The dynamical mass of the TDG can be estimated as  (Monreal Ibero et al. \citeyear{mon07}):

$$ \rm M_{dyn} (\msun) = {\it constant}  \times 10^6 ~\frac{R_{eff}}{\rm kpc} ~(\frac{\sigma}{\rm \kms})^2$$

The $constant$ value ranges between 1.4 and 2.2. Following  Monreal Ibero et al. (\citeyear{mon07}) we adopt a value of 2.09. This calculation is highly uncertain due to the  uncertainty on the exact mass distribution of the object and specially on the velocity dispersion $\sigma$=FWHM/2.35. In spite of this, an estimation of the ranges allowed can be useful to support or discard the TDG scenario. The FWHM$_{\rm CO}$=165$\pm$85 \kms of J0025B\footnote{This value is larger than 80$\pm$40 \kms~ quoted in \S3.2 for the red fainter component of the double horned CO profile. The reason is that    the total CO emission associated with J0025B has been considered, rather than only  the inner 2.3"$\times$2.3" {\it pixel} used to build the 1-dimensional spectrum discussed in that section.} ~is not a reliable tracer of the dynamical mass, since, as we have seen, it is contaminated by emission form other regions. On the other hand, the optical emission lines, which are much less affected by this problem, have FWHM(H$\beta$)$\leq$140 \kms, FWHM([OIII]$\lambda$5007)=330$\pm$10 \kms (\cite{vm11a})  and  FWHM([OI]$\lambda$6300)=140$\pm$20 \kms. Such discrepancy for [OIII] might indicate that it is broadened by some turbulent mechanism (e.g. winds).  It is reasonable to assume FWHM$\sim$140 \kms or $\sigma\sim$60 \kms  (notice that  this value also locates the system  with other TDGs in the $R_{eft}$ vs. $\sigma$ plane; e.g. Monreal Ibero et al. \citeyear{mon07}). In such case, M$_{\rm dyn}\sim$5$\times$10$^9$ \msun. Dynamical masses  as large as this have been measured for other TDG (Monreal Ibero et al. \citeyear{mon07}, Bournaud \& Duc \citeyear{bou06}). At the lowest extreme, even unexpectedly small $\sigma$ values (e.g. 10 \kms) would result in M$_{\rm dyn}>$10$^8$ \msun, still typical of TDGs. Thus, in spite of the uncertainties, we confirm that the dynamical mass of the optical tidal feature associated with J0025B centroid is in the range of TDGs.

The formation of such massive self-gravitating condensations of matter in the outer regions of tidal tails is allowed by the simulations,
 without strong constraints
on the galactic encounter parameters (Bournaud \& Duc \citeyear{bou04}). These condensations can survive longer ($>$1 Gyr) than the inner regions of the tails.  According to these simulations, only the objects
formed at the tip of tidal tails can become long-lived, massive, dwarf satellite galaxies. Whether this will be the case  depends on its stability against its own internal motions and against tidal forces from the parent galaxies.

 \begin{figure}
\includegraphics{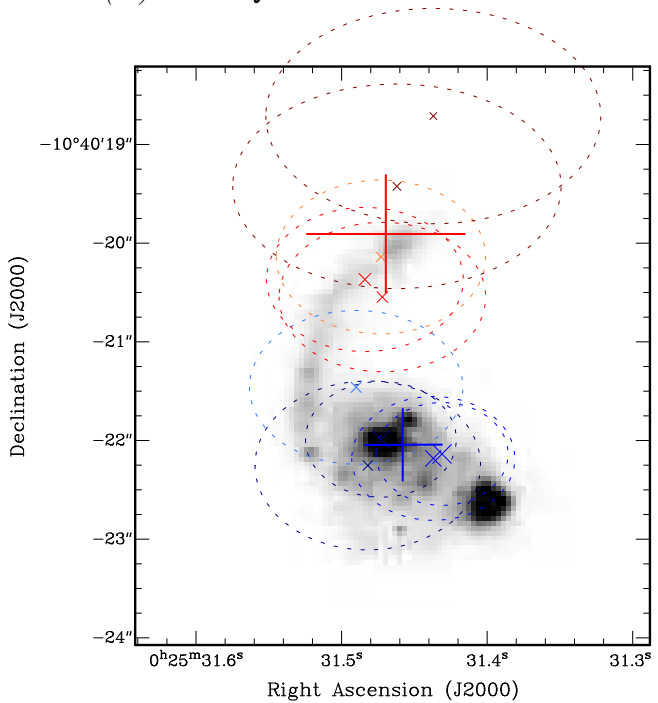}
\vspace{3.5in}
\caption{Overview of the location of the CO(1-0) peak emission against features seen in the HST image (grey-scale). The large `+' signs indicate the location of the peak of the CO(1-0) emission in J0025A (blue) and J0025B (red) derived from the total intensity image of Fig.\,3 (see text for details). The region covered by the size of the `+' symbol indicates the uncertainty in the exact location as a result of uncertainty in the fitting of the CO(1-0) peak combined with astrometric uncertainty (\S2). The crosses indicate the location of the peak of the CO(1-0) emission in the various channel maps of Fig.\,3 (dark blue $\rightarrow$ dark red for lowest $\rightarrow$ highest velocities). The size of the cross scales with the significance level at which the CO(1-0) is detected, while the dashed ellipses indicate the uncertainty in the exact location (from the CO(1-0) fitting combined with astrometric errors). A color version of this figure can be seen in the electronic form.}
\end{figure}

\section{Summary and conclusions}

We have characterized the amount,  spatial distribution and kinematics of the molecular gas in the optically selected type 2 quasar SDSS J0025-10 at $z=$0.30, which is a merging system with a double composite nucleus (AGN+HII). For this, we have used the CO(1-0) transition based on data obtained with the Australia Telescope Compact Array (ATCA).  This is one of the  scarce examples of quasar host galaxies  where the CO emission has been resolved  spatially at any redshift.

  The system contains  \mh2=(6$\pm$1)$\times$10$^9$ \msun, which is distributed in two main reservoirs separated by $\sim$9 kpc. $\sim$60\% of the gas is in the central region, preferentially associated with the QSO nucleus and/or the intermediate region between the two nuclei. The other 40\% 
 is associated with the northern tidal tail and is therefore unsettled.

 Based on WISE and IRAS photometry, we constrain the infrared luminosity  \lir=(1.1$\pm$0.3)$\times$10$^{12}$ \lsun~and show that it is dominated by a starburst, rather than the AGN. SDSS J0025-10 is an analogue of local ($z\la$0.05) luminous LIRGs and ULIRGs. The clear morphological evidence for an ongoing major merger, the large molecular gas content,  the star forming rate SFR=190$\pm$52 \lsun~ \msun$^{-1}$, the star formation efficiency   SFE=183$\pm$62 \lsun~ \msun$^{-1}$ and the gas exhaustion time scale $\tau_{\rm SF}$ = 32$\pm$11 Myr are   all in the range  of these local systems.

SDSS J0025-10 is the result of  an ongoing major merger of two gas rich progenitors. This, together with the high infrared luminosity  dominated by a starburst,  the massive reservoir of molecular gas and the quasar activity  are all properties consistent with a transition phase in the (U)LIRG-optical QSO evolutionary scenario. The relatively large  nuclear separation ($\sim$5 kpc) and the existence of large amounts of unsettled molecular gas (probably surviving gas from the merger), moreover suggest that we are observing the system during a particular transient phase, prior to more advanced mergers where the nuclei have already coalesced. 
  
 We propose that at least part of the extended, unsettled reservoir of molecular gas is associated with a tidal dwarf galaxy  at the tip of the northern tidal tail identified in the HST image. The formation of tidal dwarf galaxies in the outer regions of tidal tails is predicted by simulations of colliding galaxies.
  
 \section*{Acknowledgments}
 Thanks to an anonymous referee for  useful comments
that helped improve the paper substantially. Thanks to Luis Colina for useful discussions and suggestions on the manuscript. 

This work has been funded with support from the Spanish
former Ministerio de Ciencia e Innovaci\'on through the
grant AYA2010-15081. ATCA is funded by the Commonwealth of Australia for operation as a National Facility managed by CSIRO.

\end{document}